\documentclass[reqno]{amsart}

\usepackage{amsfonts}
\usepackage{bezier}
\usepackage{euscript}
\usepackage{amsfonts,amssymb,amsmath,amsbsy,amsthm,amscd}
\usepackage{xcolor}
\usepackage{graphicx}
\usepackage{color}
\usepackage[T2A]{fontenc}
\usepackage[utf8]{inputenc}
\usepackage[english]{babel}   

\newtheorem{Theorem}{Theorem}
\newtheorem{Lemma}{Lemma}
\newtheorem{cor}{Corollary}
\newtheorem{rem}{Remark}

\begin{document}

\title[The fundamental solution for a jump-diffusion Ornstein-Uhlenbeck
 ]{
The fundamental solution of the master equation for a jump-diffusion Ornstein-Uhlenbeck process
}


\author{ Olga S. Rozanova*, Nikolai A. Krutov}

\address{ Mathematics and Mechanics Department, Lomonosov Moscow State University, Leninskie Gory,
Moscow, 119991, Russian Federation}
\email{rozanova@mech.math.msu.su}



\begin{abstract}
An integro-differential equation for the probability density of the generalized stochastic Ornstein-Uhlenbeck process with jump diffusion is considered for a special case of the Laplacian distribution of jumps. It is shown that for a certain ratio between the intensity of jumps and the speed of reversion, the fundamental solution can be found explicitly, as a finite sum. Alternatively, the fundamental solution can be  represented as converging power series.  The properties of this solution are investigated. The fundamental solution makes it possible to obtain explicit formulas for the density at each instant of time, which is important, for example, for testing numerical methods.
\end{abstract}

\maketitle

\section{Introduction}

Models using stochastic dynamics have natural applications in various areas of physics, biology, and financial mathematics. In recent decades, it has become clear that many phenomena cannot be explained by adding only standard Wiener processes to deterministic models, it is necessary to consider models that take into account differently distributed jumps, that is, use non-Gaussian stochastic models. For example, let us mention some works where non-Gaussian models are used in  physics of metals \cite{Kogan}, \cite{Billings}, in the study of neural networks \cite{Wang} and genome behavior \cite{Chen}, \cite{Marko}, in weather forecasting \cite{Yang}, and in financial mathematics \cite{Cont}.


Although models using non-Gaussian stochastic dynamics are quite diverse, their probability density necessarily obeys some generalized Kolmogorov-Fokker-Planck equation containing a non-local (integral) term. Such equations are sometimes called the Kolmogorov-Feller equations. Many mathematical works  study the existence of density and its smoothness for various types of non-Gaussian processes, the properties of transition probability \cite{Picard}, \cite{Priola}, \cite{Knopova}, \cite{Kuhn}, \cite{Peszat}. However, in practice, when solving such equations, one usually has to use numerical methods \cite{Borzi}.

In this paper, we consider the simplest generalization of the Ornstein-Uhlenbeck process to the case of jump diffusion. Such processes have traditional applications to active particle dynamics \cite{Denisov}, as well as to  modeling of interest rates in financial mathematics \cite{Maller}.

Previously, it was known that the assumption of a connection between the force acting on the particle and the properties of the kernel of the jump process helps to construct an exact solution of the corresponding stationary Kolmogorov-Feller equation, that is, to study the large time density distribution \cite{Denisov}, \cite{Rudenko}.

However, it was not noticed earlier that in some cases the assumption of a connection between the intensity of jumps and the speed of reversion allows one to obtain an explicit formula for the fundamental solution, and hence, to obtain an integral formula for the dynamics of an arbitrary initial density as a convolution. Moreover, for some initial densities (e.g. Gaussian or piecewise constant density), one can obtain an explicit formula describing the dynamics of the density at all times. This is valuable, in particular, for testing numerical algorithms.

The properties of fundamental solutions for integro-differential equations have been studied in previous works. For example, in \cite{Piatnitski} the asymptotics of the fundamental solution was constructed depending on the properties of the kernel of the process describing the jumps. It was also noted  that in the case of pure jumps (without diffusion), the fundamental solution always contains a singular component.

In this paper, a fundamental solution of the Kolmogorov-Feller equation is constructed for the case of a Laplacian distribution of jumps. In the general case, the fundamental solution can be written as a series; however, with a countable number of dependencies between the return force and the intensity of the jump process, this series reduces to a finite sum. These exact formulas make it possible, in particular, to study in detail the smoothness of the regular part of the fundamental solution, the asymptotic behavior of its tails, and its behavior in the limit at large times.

It should be noted that fundamental solutions are known for various evolutionary integro-differential equations, including the density equation in the case of anomalous diffusion, for example, \cite{Mainardi}, \cite{Luchko}. As a rule, they have the form of an integral transform or can be written as a series of special functions, but with a certain combination of parameters, fundamental solutions can be written in a closed form \cite{Ferreira}.

In the case of the density equation of the generalized Ornstein-Uhlenbeck process, more complex models can be considered, for example, with other jump distributions. However, it seems that with further modifications, it is impossible to obtain such results by practically elementary methods, as in this work.


\section{Probability density for a jump diffusion model}

 Let $X_t$ be  a stochastic process with  dynamics given by
\begin{equation}\label{X}
dX_s = (B-\beta\,X_s)\, d s + \sigma\, d W_s + \lambda d \Gamma_s, \quad X_0=x_0,
\end{equation}
 $x_0\in \mathbb R$  is a point in the space of states,
  $0\le s\le T$,
$W_s\in \mathbb R$ is a standard  Brownian motion, $\Gamma_s\in \mathbb R$ is the compound Poisson process
with the generator $\mathcal L f=\int\limits_{\mathbb R} (f(x+y)-f(x)) p(y) dy$, where
$p(z)$ is a probability density of jumps, $\int\limits_{{\mathbb R}}\,p(z) dz=1$,
$\beta>0$, $B$, $\sigma\ge 0$,  $\lambda\ge 0$
 are constants.

 If $\lambda=0$, process \eqref{X} is a standard Ornstein-Uhlenbeck process.


Let $P(t,x)\ge 0$ be the probability density of $X_s$.
 We consider a particular case of Laplace distribution with the kernel  $p(z)=\frac{k}{2 }\,{{\rm e}^{-k
  \left| z\right| }}$, $k>0$.

The Kolmogorov-Feller equation for the function $P=P(t,x)$, $x\in\mathbb R$, $t\ge 0$ has the form (e.g. \cite{Schuss})
\begin{eqnarray}\label{FPF}
&&{\frac {\partial }{\partial t}}P \left( t,x \right)  + {\frac {\partial }{\partial x}}
 \left(\left( B-\beta\,x \right)\,P \left( t,x \right)\right) -\frac12\,{\sigma}^{2}{\frac {\partial ^{2}}{
\partial {x}^{2}}}P \left( t,x \right) -\\
&&{\lambda}\, \left(\frac12\,{
k}\,\int _{-\infty }^{\infty }\!P \left( t,x-z \right) {{\rm e}^{-k
 \left| z \right| }}{dz}-P \left( t,x \right)  \right)=0.\nonumber
 \end{eqnarray}

 The fundamental solution $\mathcal E(t,x,y)$ is  the solution to the Cauchy problem \eqref{FPF} with the initial data
 \begin{equation*}
  P|_{t=0}=\delta(x-y).
 \end{equation*}

If $\mathcal E(t,x,y)$ is known,
then the solution of the Cauchy problem with any other integrable initial data $P|_{t=0}=\phi(x)$, $\int\limits_{\mathbb R} \phi \, dx=1$,
can be found as
 \begin{equation}\label{conv}
P(t,x)=\int _{-\infty }^{\infty }\!{\mathcal E} \left( t,x,y \right) \phi(y) {dy}.
 \end{equation}

For the standard Ornstein-Uhlenbeck process ($\lambda=0$) the fundamental solution is well known, see, e.g. \cite{Gardiner}.

 The Fourier transform $\hat {\mathcal E}(t,w,y)$ solves the following problem:
 \begin{eqnarray}\label{FT}
 &&{\frac {\partial }{\partial t}}\hat  {\mathcal E}   + \beta w {\frac {\partial }{\partial w}}
 \,\hat  {\mathcal E}+ \\&&  \frac{1}{2 (w^2+k^2)}   \left[ \sigma^2 w^4
+2\,iB{w}^{3}+ \left(\sigma^{2}{k}^{2}+2\,
{\lambda} \right) {w}^{2}+2\,iBw{k}^{2}\right]\,\hat  {\mathcal E} =0,\nonumber\\&& \quad \hat {\mathcal E}|_{t=0}=e^{-i w y}.\nonumber
 \end{eqnarray}
The solution of \eqref{FT} can be  found in the standard way,
\begin{equation}\label{FT_sol}
\hat  {\mathcal E}(t,w,y)= \left(1+\frac{({ e}^{-2\,\beta\,t}-1)\, {w}^{2}}{ {k}^{2}+{w}^{2}} \right) ^{\frac
{\lambda}{2 \beta}}{{ e}^{{\frac {w \left( {{\rm e}^{-\beta\,
t}}-1 \right)  \left( w {\sigma}^{2}({{ e}^{-\beta\,t}}+1)+4\,iB \right) }{4\beta}}-i w e^{-\beta t}y}}.
\end{equation}
Further, we denote $\alpha=\frac
{\lambda}{2 \beta}\ge 0$.

We note that the Fourier transform of the fundamental solution  can be found analytically for many models, except those considered here, for more complex kernels of the jump distribution. However, the inverse Fourier transform does not lead to an explicit formula, so we can only be satisfied with the integral representation of the solution.

  \section{Fundamental solution and its properties}

Notice that the first multiplier of \eqref{FT_sol} can be expanded into an absolutely and uniformly convergent  power series (for all $w\in \mathbb C$, $t>0$) as
\begin{equation*}\left(1+\Psi(t,w) \right) ^{\alpha}=\sum\limits_{n=0}^\infty  \left(
\begin{array}{c}
 \alpha \\
  n \\
  \end{array}
  \right)\,\Psi^n(t,w),\qquad n\in\mathbb N \cup\{0\},
\end{equation*}
where
\begin{eqnarray*}
 \Psi=\frac{({ e}^{-2\,\beta\,t}-1)\, {w}^{2}}{ {k}^{2}+{w}^{2}},
\end{eqnarray*}
$\left(
\begin{array}{c}
 \alpha \\
  n \\
  \end{array}\right)$
  is the (generalized) binomial coefficient. Thus,
  \begin{equation}\label{FT_sum}
\hat  {\mathcal E}(t,w,y)= {{ e}^{{\frac {w \left( {{\rm e}^{-\beta\,
t}}-1 \right)  \left( w {\sigma}^{2}({{ e}^{-\beta\,t}}+1)+4\,iB \right) }{4\beta}}-i w e^{-\beta t}y}}\, \sum\limits_{j=0}^\infty  \left(
\begin{array}{c}
 \alpha \\
  j \\
  \end{array}
  \right)\,\Psi^j(t,w).
\end{equation}

Let us introduce new functions for $n\in\mathbb N \cup\{0\}$:
\begin{eqnarray}\label{Fn}
 && F_n(t,w)=\frac{\exp( A_1 i w+A_2 w^2)}{(k^2+w^2)^n},\nonumber\\ && A_1=-\frac{B}{\beta}(1-e^{-\beta t}), \quad  A_2=-\frac{\sigma^2}{4\beta}(1-e^{-2\beta t}).\label{A12}
 \end{eqnarray}
  We denote $[F(t,w)]$  the inverse Fourier transform with respect to $w$.
 Since  the multiplication by
$\exp( A_1 i w)$ in $[F_n(t,w)]$   only leads to the replacement of the argument $x$ by $x+A_1$,
we can perform computations in the shifted variables.

Further we use the notation $\bar x=x-y e^{-\beta t}+A_1$. The operator $D^{(k)}_{\bar x}$ means differentiation of order $k$, $k\in {0}\cup \mathbb N$ with respect to $\bar x$.

\begin{Lemma}\label{L1} For any $\alpha>0$ the fundamental solution has the following formal representation:
\begin{equation}\label{FT_sum2}
 {\mathcal E} (t,x,y)= {\mathcal E}(t,{\bar x})=  \sum\limits_{j=0}^\infty  \left(
\begin{array}{c}
 \alpha \\
  j \\
  \end{array}
  \right)\,\left(1- {{\rm e}^{-2\beta\,
t}} \right)^j\,D^{(2j)}_{\bar x}[F_j(t,w)](t,{\bar x}).
\end{equation}
\end{Lemma}

\proof
We have
  \begin{eqnarray*}\label{Psi_n}
&&  \left[{{ e}^{{\frac {w \left( {{\rm e}^{-\beta\,
t}}-1 \right)  \left( w {\sigma}^{2}({{ e}^{-\beta\,t}}+1)+4\,iB \right) }{4\beta}}-i w e^{-\beta t}y}}\, \Psi^j(t,w)\right](t,{\bar x})=\\
&&\left( {{\rm e}^{-2\beta\,
t}}-1 \right)^j [w^{2j}\,F_j(t,w)](t,{\bar x})= \left(1- {{\rm e}^{-2\beta\,
t}} \right)^j\,D^{(2j)}_{{\bar x}}[F_j(t,w)](t,{\bar x}).
  \end{eqnarray*}
 therefore \eqref{FT_sum} implies \eqref{FT_sum2}.
$\Box$


 Then we want to show that for some relations between $\lambda$ and $\beta$  one can to obtain  the fundamental solution as a finite sum.
\begin{Lemma}\label{L1}  
Assume $\alpha=n$ or
$\lambda=2n\beta,$ $n\in \mathbb N.$
 Then
 \begin{equation}\label{FSol}
 {\mathcal E} (t,x,y)=
 {\mathcal E} (t,{\bar x})=\sum\limits_{j=0}^n (-1)^j\left(
\begin{array}{c}
 n \\
  j \\
  \end{array}\right) k^{2(n-j)} D_{{\bar x}}^{(2j)}[ F_n(t,w)](t,{\bar x}) e^{-2\beta t j}.
 \end{equation}
\end{Lemma}

 \proof The representation \eqref{FSol} is a particular case of \eqref{FT_sum2}, therefore \eqref{FT_sum2} can be transformed to \eqref{FSol} taking into account properties of $F_n$. However, it is easier (and more convenient for us) to obtain it directly  from \eqref{FT_sol} by Fourier transform  if we rewrite  \eqref{FT_sum} as
 \begin{eqnarray*}
\hat  {\mathcal E}(t,w)=
(k^2+e^{-2kt}w^2)^n F_n(t,w)
\end{eqnarray*}
and then apply Newton's binomial formula. $\Box$

\subsection{Properties of $[F_n]$}

We consider two cases: $\sigma=0$, which corresponds to $A_2=0$, and $\sigma \ne 0$, where $A_2<0$ for $t>0$.
In order not to clutter up the notation, in this subsection we  write $x$ instead of ${\bar x}$.

\begin{Lemma}\label{L2}
In the case of $\sigma=0$, the functions $[F_n](x)$ solve  the linear differential equation with constant coefficients
(with respect to $x$),
\begin{equation}\label{Fn_eq}
 a_{n-1} D^{n-1}_x [F_n]+\dots + a_{0} [F_n]=x^{n-1} [F_1], \quad n\in \mathbb N,
 \end{equation}
 satisfying the condition 
 $D^{j}_x[F_n]\to 0$ as $|x|\to \infty$, $j=0,...,n-1$,
 where the coefficients $a_j$ correspond to the powers of $w$ in the numerator of the expression
\begin{equation*}\label{a_coeff}
D^{n-1}_x \left(\frac{1}{k^2+w^2}\right)= \frac{a_{n-1} w^{n-1}+\dots + a_{0}}{(k^2+w^2)^n}.
\end{equation*}
\end{Lemma}

\proof
On the one hand
\begin{equation*}\label{FD1}
[D^{n-1}_x F_1]  = (-i)^{n-1} x^{n-1} [F_1],
\end{equation*}
on the other hand
\begin{equation*}\label{FD2}
[D^{n-1}_x F_1]  = a_{n-1}(-i)^{n-1} D^{n-1}_x [F_n]+...+a_0 [F_n],
\end{equation*}
then after  dividing by $(-i)^{n-1}$ we get \eqref{Fn_eq}. $\square$

\begin{Lemma}\label{L3}
In the case of $\sigma\ne 0$, the functions $[F_n](t,x)$ solve  the linear differential equation with time-dependent coefficients
\begin{equation}\label{Fn_eqs}
 a_{3(n-1)}D^{3(n-1)}_x [F_n]+\dots + a_{0} [F_n]=\,x^{n-1} [F_1], \quad n\in \mathbb N,
 \end{equation}
 satisfying the condition
 $D^{j}_x[F_n]\to 0$ as $|x|\to \infty$, $j=0,...,3n-4$,
 where the coefficients $a_j(t)$ correspond to the powers of $w$ in the numerator of
\begin{equation*}\label{a_coeff}
D^{n-1}_x \left(\frac{e^{A_2 w^2}}{k^2+w^2}\right)= \,\frac{a_{3(n-1)} w^{3(n-1)}+\dots + a_{0}}{(k^2+w^2)^n} e^{A_2 w^2}.
\end{equation*}
\end{Lemma}

\proof Similar to the proof of Lemma \ref{L2}
\begin{equation*}\label{FD1s}
[D^{n-1}_x F_1]  = i^{n-1} x^{n-1} [F_1]=
  a_{3(n-1)}(-i)^{3(n-1)} D^{3(n-1)}_x [F_n]+...+a_0 [F_n],
\end{equation*}
after dividing by $i^n$ we get \eqref{Fn_eqs}. $\square$

Below we  use the standard notation ${\rm Erf}(x)=\frac{2}{\sqrt{\pi}}\int\limits_{0}^{x} e^{-\xi^2} d\xi$.

\begin{Lemma}\label{L4}
For $\sigma=0,$ we have
\begin{equation}\label{F10}
 [F_1](x)= \frac{1}{2k} e^{-k|x|},
 \end{equation}
 for $\sigma\ne 0$ we have
 \begin{eqnarray}\label{F1s}
 &&[F_1](t,x)=\frac{e^{-k^2A_2}}{2} \left(\cosh k x+\frac{1}{2k} \left[{\rm Erf}\left(\frac{x}{2\sqrt{-A_2}}-k\sqrt{-A_2}\right) e^{kx}+\right.\right.\\ &&\left.\left.{\rm Erf}\left(-\frac{x}{2\sqrt{-A_2}}-k\sqrt{-A_2}\right) e^{-kx}\right]\right),\quad A_2=A_2(t),\nonumber
 \end{eqnarray}
 moreover,
 \begin{equation}\label{F1ss}
D^2_x [F_1](t,x) - k^2 [F_1] (t,x) = -\frac{1}{2\sqrt{-\pi A_2(t)}} e^{\frac{x^2}{4 A_2(t)}}.
 \end{equation}
 \end{Lemma}

 The {\it proof} is a direct computation. $A_2$ is given in \eqref{A12}. $\square$

 \begin{cor}\label{cor1}
1. For $\sigma=0$
  $$[F_n](x) \in C^{2n-2}({\mathbb R}),$$
$D^{2n-1}_x [F_n](x)$ has a discontinuity of the first kind at zero;

 2. For $\sigma>0$
 $$[F_n](t,x)\in C^\infty({\mathbb R})$$
 as a function of $x$.
 \end{cor}

\proof 1. According to \eqref{Fn_eq} $[F_n]$ is the solution of the  inhomogeneous linear differential equation with constant coefficients with a right-hand side having smoothness $C^{n-1}$ with respect to $x$.  Therefore $[F_n]$ has the smoothness $n-1$ units higher than the right side, i.e. $C^{2n-2}$. The term with the highest derivative, $D^n_x[F_n]$, has the same smoothness properties as the right hand side $x^{n-1} [F_1]$, therefore
$D^{2n-1}_x [F_n](x)$ has a discontinuity of the first kind at zero. Since the right side does not depend on $t$, the size of the jump also does not depend on $t$.

2. According to \eqref{F1ss} $[F_1]$ is the solution of the  inhomogeneous linear differential equation  (in $x$) with constant coefficients for any fixed $t$  with an infinitely differentiable right hand side, so it belongs to $C^\infty({\mathbb R})$. Thus, the right hand side of \eqref{Fn_eqs} belongs to $C^\infty({\mathbb R})$. Since \eqref{Fn_eqs} is  a  nonhomogeneous linear equation with constant coefficients, then
$[F_n](t,x)\in C^\infty({\mathbb R})$ with respect to $x$.
 $\square$

 \begin{cor}\label{cor2}
1. For $\sigma=0$
  $$[F_n](x) \sim  {\rm const} \, x^{n-1}\,e^{-k|x|}, \quad |x|\to \infty;$$

 2. For $\sigma>0$
 $$[F_n](t, x)\sim   \frac{ {\rm const}}{\sqrt{- A_2(t)}} \, x^{n-1}\, e^{\frac{x^2}{4 A_2(t)}}, \quad |x|\to \infty, \quad t>0.$$
 \end{cor}

 The {\it proof} follows from Lemmas 3 and 4 and the formula of representation of solution for a nonhomogeneous linear equation. Recall that $A_2<0$ for $t>0$. $\square$

\subsection{Main results}

Let us summarize our results.
\begin{Theorem}\label{T1}

For any $\beta>0$, $\lambda\ge 0$
 the fundamental solution ${\mathcal E}(t,x,y)$ of equation \eqref{FPF}  can be found as \eqref{FT_sum2}, where $\alpha=\frac{\lambda}{2 \beta}$.

  \begin{itemize}

\item For $\sigma=0$ the quantities $[F_n(t,w)]$ can be found as solutions of  equations \eqref{Fn_eq}, with \eqref{F10} on the right side,  with the argument $ x$ replaced by $\bar x=x-y e^{-\beta t}+A_1$.  The fundamental solution  ${\mathcal E}(t,x,y)$  (given as \eqref{FSol}) is a sum of the singular component ${\mathcal E}_s(t, x, y)$ and the regular component ${\mathcal E}_r(t,x,y)$, where
    \begin{eqnarray}\label{q1}
 {\mathcal E}_s(t,x,y) =   q(t)\,\delta\left(x-y e^{-\beta t}-\frac{B}{\beta}(1-e^{-\beta t})\right),\qquad  q(t)={{\rm e}^{-2\alpha \beta
t}},
 \end{eqnarray}
    and ${\mathcal E}_r(t,x,y)={\mathcal E}(t,x,y)-{\mathcal E}_s(t, x, y) =  {\mathcal E}_r(t,\bar x)$. The series for the regular part
     converges
    \begin{itemize}
    \item for $\alpha>\frac12$ at   any point $\bar x\in\mathbb R$,
    \item  for $0<\alpha\le\frac12$ at  any point $\bar x\in\mathbb R\setminus{0}$; at the point $\bar x=0$ the function ${\mathcal E}_r(t,x,y)$ has an integrable singularity,
      \item for $\alpha=0$ the fundamental solution contains the singular component  ${\mathcal E}_s(t, x, y)$ only, i.e. ${\mathcal E}_r(t, x, y)=0$.
\end{itemize}
  \item   For $\sigma>0$ $[F_n(t,w)]$ can be found as solutions of  equations
 \eqref{Fn_eqs}, with \eqref{F1s} on the right side, with the argument $x$ replaced by $\bar x$.
Series \eqref{FT_sum2} converges for any $\bar x\in\mathbb R$,
${\mathcal E}(t,x,y)$ is a $C^\infty$  function with respect to all variables.

\end{itemize}

\end{Theorem}

\proof

For the case $\sigma=0$ the amplitude of the delta-function  can be found from \eqref{Fn_eq}, taking into account the properties
of \eqref{F10}. First of all, we notice that  $[F_0(t,w)]=\delta(\bar x)$. Further, to find $D^{(2j)}_{\bar x}[F_j(t,w)]$ we change the index $n$ to $j$ in \eqref{Fn_eq} and apply the operator $D^{j+1}_{\bar x}$ to both sides of \eqref{Fn_eq} for $j\in \mathbb N$. The singular component arises in the left hand side from the first term only. On the right hand side we have
\begin{eqnarray}\label{ind}
 J_j\equiv D_{\bar x}^{j+1}({\bar x}^{j-1}[F_1])= \,- j!\, \delta({\bar x})+R_j({\bar x}), \quad j\in \mathbb N,
  \end{eqnarray}
where $R_j({\bar x})$ does not contain a singular component ($R_j({\bar x})\in C(\mathbb R)$, but has no derivative at ${\bar x} =0$).
Let us prove this formula by induction. Indeed, it is easy to check that
\begin{eqnarray*}
D_{\bar x} [F_1] (\bar x)= - k \,{\rm sign} \,\bar x \, [F_1](\bar x),  \qquad D_{\bar x}^2 [F_1](\bar x)= - \delta(\bar x) + k^2 \, [F_1](\bar x), \end{eqnarray*}
and the term $D_{\bar x}^{\gamma_1} ({\bar x}^{\gamma_2} [F_1]({\bar x}))$, $\gamma_1, \,\gamma_2 \in {0} \cup \mathbb N$, $\gamma_1\le \gamma_2$, does not contain a singular component. We have
\begin{eqnarray*}
J_1= - \delta + k^2 \, [F_1], \qquad J_2= - 2\,\delta + (3- k\, {\bar x} \,{\rm sign}\,{\bar x} ) k^2 \, [F_1],
\end{eqnarray*}
 so \eqref{ind} is valid  for $j=1$, $j=2$. Assume $j>2$. Then
\begin{eqnarray*}
&& D_{\bar x}^{j+1}({\bar x}^{j-1}[F_1])= D_{\bar x}^j ((j-1) {\bar x}^{j-2}[F_1]+{\bar x}^{j-1} D_{\bar x} [F_1])=\\&& - (j-1) (j-1)! \, \delta({\bar x})+ D_{\bar x}^j({\bar x}^{j-1} D_{\bar x} [F_1])=\\
&&- (j-1) (j-1)! \,\delta({\bar x}) + D_{\bar x}^{j-1}((j-1) {\bar x}^{j-2} D_{\bar x} [F_1] +  {\bar x}^{j-1} D_{\bar x}^2 [F_1])=
\\
&&- (j-1) (j-1)! \,\delta({\bar x}) + D_{\bar x}^{j-2}((j-1)(j-2) {\bar x}^{j-3} D_{\bar x} [F_1] + \\&& (j-1) {\bar x}^{j-2} D_{\bar x}^2 [F_1])+ R_j({\bar x})=\dots =\\&&- (j-1) (j-1)! \, \delta({\bar x}) -
(j-1)! \, D_{\bar x}^2 [F_1] + R_j({\bar x}) = -j!  \,\delta(x) + R_j ({\bar x}).
\end{eqnarray*}
We  use the fact that  $ \bar x^\beta D_{\bar x}^2 [F_1]({\bar x})= - \bar x^\beta \delta(\bar x) + k^2 \bar x^\beta [F_1]({\bar x}) = k^2 \bar x^\beta [F_1]({\bar x}) $ for $\beta>0$. In these calculations,   $R_j$ are different non-singular terms.

Thus, for any $j\in \mathbb N$, the amplitude $q(t)$ of the singular component is
$\displaystyle -\frac{j!}{a_{j-1}}$. Since $a_{j-1}=(-1)^{j-1} j!$, then
substitution into \eqref{FT_sum2} gives
\begin{eqnarray*}
 q(t)=
  1+\sum\limits_{j=1}^\infty  \left(
\begin{array}{c}
 \alpha \\
  j \\
  \end{array}
  \right)\,{(-1)^{j} \left(1- {{\rm e}^{-2\beta\,
t}} \right)^j}=
(1- (1- {\rm e}^{-2\beta
t}))^\alpha
= {{\rm e}^{-2\alpha \beta
t}} ,
    \end{eqnarray*}
which results in \eqref{q1}.

\bigskip

The convergence of ${\mathcal E}_r(t,{\bar x})$ for $\sigma=0$ is difficult to obtain from the series itself, therefore we analyse
 $\hat {\mathcal E}$ in \eqref{FT_sol} and study the convergence of the integral
 \begin{equation}\label{int}
 \int\limits_{\mathbb R} \hat {\mathcal E} (t, w, y) \, e^{i w x} \,dw
\end{equation}
 for the inverse Fourier transform. It is sufficient to study the case $y=B=0$, i.e. $\bar x =x$. We start with the limit case $t\to\infty$ and denote  $ \hat {\mathcal E}_\infty (w) = \lim\limits_{t\to\infty} \hat {\mathcal E}$.  It is easy to see that $ \hat {\mathcal E}_\infty (w)\to 0$ as $|w|\to \infty$.  For $\alpha>\frac12$
 \begin{equation*}
 \hat {\mathcal E}_\infty (w)= \left(\frac{{k}^{2}}{ {k}^{2}+{w}^{2}} \right) ^{\alpha}
 \end{equation*}
 belongs to $L_1(\mathbb R)$ and the inverse Fourier transform exists for all $x\in \mathbb R$. If $\alpha\le \frac12$, then  the integral \eqref{int} diverges for $x=0$.

  However, for $x\ne 0$
  \begin{equation*}
 \int\limits_{\mathbb R} \hat {\mathcal E}_\infty ( w ) \, e^{i w x} \,dw = - \frac{1}{i x}\,\int\limits_{\mathbb R} D_w (\hat {\mathcal E}_\infty ( w )) \, e^{i w x} \,dw = -\frac{2}{ x} \, \int\limits_0^\infty D_w \hat {\mathcal E}_\infty ( w ) \, \sin {w x} \,dw,
\end{equation*}
 the latter integral converges. Moreover, it can be readily shown that
 \begin{equation*}
 \int\limits_{\mathbb R}\int\limits_{\mathbb R} \hat {\mathcal E}_\infty ( w ) \, e^{i w x} \,dw \, dx = 4 \lim\limits_{x\to\infty}\int\limits_0^\infty \frac{\sin w x}{w} \hat {\mathcal E}_\infty ( w )  \,dw\le 2\pi,
\end{equation*}
 therefore the integral converges. Here we used the properties of the Dirichlet integral and the boundedness of $\hat {\mathcal E}_\infty$. The value of the limit is necessarily equal to 1,  since the integral of the density $P(t,x)$ over the axis $\mathbb R$ is preserved provided it exists. Thus,  the singularity at $x=0$ is integrable.

 For any finite $t>0$ the integral \eqref{int} diverges, which means the presence of the singular component. However, if we want to study the regular component, we have to consider the integral
 \begin{equation*}
 \int\limits_{\mathbb R} \, (\hat {\mathcal E} (t, w, y) - \hat {\mathcal E}_s (t, w, y))\, e^{i w x} \,dw.
\end{equation*}
 As follows from \eqref{q1}, for $y=B=0$
  \begin{equation*}
  \hat {\mathcal E} (t, w, y) - \hat {\mathcal E}_s (t, w, y)= \left(1+\frac{({ e}^{-2\,\beta\,t}-1)\, {w}^{2}}{ {k}^{2}+{w}^{2}} \right) ^{\alpha} - {\rm e}^{-2 \alpha \beta t }\to 0, \quad |w|\to \infty,
  \end{equation*}
therefore to obtain the statement of the theorem, it is enough to repeat the same reasoning as in the case $t\to \infty$.

For $\alpha=0$ the amplitude of the singular component ${\mathcal E}_s$ is constant, $q(t)=1$, which implies ${\mathcal E}_r=0$.

\bigskip

2. For $\sigma>0$ the convergence of ${\mathcal E}(t,{\bar x})$  for any ${\bar x}\in\mathbb R$ and  $t>0$ follows from the existence of the inverse Fourier transform of $\hat {\mathcal E}$ given as \eqref{FT_sol} due to the presence of the exponential multiplier.
$\square$

\bigskip
\begin{rem}
For $\alpha\le \frac12$ the regular part  ${\mathcal E}_r$ is not continuous, it has an integrable singularity at $\bar x=0$. An explicit representation of the stationary probability density ${\mathcal E}_{st}$, i.e. the limit of ${\mathcal E}_r$ as $t\to\infty$  for $\alpha=\frac12$ and  $B=0$ can be found in \cite{Rudenko}:  ${\mathcal E}_{st}=\displaystyle \frac{k}{\pi} K_0(k|x|)$, where $K_0$ is the modified Bessel function of the second kind of zero order.
 \end{rem}

For  $\alpha=n$, $ n\in \mathbb N$  the fundamental solution can be written in a closed form.

\begin{Theorem} Assume $\alpha=n$, $ n\in \mathbb N$. Then the fundamental solution ${\mathcal E}(t,x,y)$ of equation \eqref{FPF}  can be found by the explicit formula \eqref{FSol} as a finite sum.

1. For $\sigma=0$
  \begin{itemize}
    \item ${\mathcal E}(t,x,y)$ necessarily contains a singular component
  \begin{equation}\label{es}
 {\mathcal E}_s(t, x, y)=
 e^{-2\beta n t}\, \delta\left(x-y e^{-\beta t}-\frac{B}{\beta}(1-e^{-\beta t})\right);
  \end{equation}
  \item the regular part of ${\mathcal E}(t,x,y)$ (${\mathcal E}_r(t,x,y)={\mathcal E}(t,x,y)-{\mathcal E}_s(t)$)  belongs to $C^{n-1}$ with respect to $x, y$;
  \item ${\mathcal E}( t,x, y)= O (x^n \, e^{-k |x|})$, $|x| \to \infty$, $y \in \mathbb R$, $t>0$ fixed.
\end{itemize}
 2.  For $\sigma>0$
 \begin{itemize}
    \item
   ${\mathcal E}(t,x,y)$ is a function of class $C^\infty$ with respect to all variables;
   \item
   ${\mathcal E}( t,x, y)= O (x^n \, e^{-\gamma x^2})$, $|x| \to \infty$, $\gamma=\gamma( t)>0$, for any $y \in \mathbb R$, $t>0$ fixed.
   \end{itemize}
    \end{Theorem}

The {\it proof} follows  from Lemmas 1, 3 -- 5. The amplitude of the delta-function is computed as in the proof of Theorem 1.

\begin{rem} Although in the case $\sigma=0$ the regular
part of the fundamental solution ${\mathcal E}_r$ generally has no derivative at $\bar x$, for $\alpha=n$ the smoothness increases to $C ^{n-1}$.
\end{rem}

\section{Examples of solutions in a closed form}

For sufficiently large $\alpha=n$, the formula for finding the fundamental solution is quite cumbersome, although it can be easily implemented using a computer algebra package. However, for small $n$ it may well be written out. Let us study it for the case $n=1$.  For simplicity, we set $y=0$.

1. For $\sigma=0$ we have
\begin{equation}\label{E10}
{\mathcal E}(t,x)=\frac{k}{2} (1-e^{-2\beta t}) e^{-k |x-\frac{B}{\beta}(1-e^{-\beta t})|} + e^{-2 \beta t } \delta \left(x-\frac{B}{\beta}(1-e^{-\beta t})\right).
\end{equation}
Note that the amplitude of the delta function changes from unity to zero, and the regular part of the solution has a nontrivial limit at $t\to \infty$:
\begin{equation}\label{P1}
{\mathcal E}_{st}(x)=  \frac{k}{2}  e^{-k |x-\frac{B}{\beta}|}, \qquad t\to \infty.
\end{equation}

\begin{figure}[htb]
\begin{minipage}{0.4\columnwidth}
\includegraphics[scale=0.35]{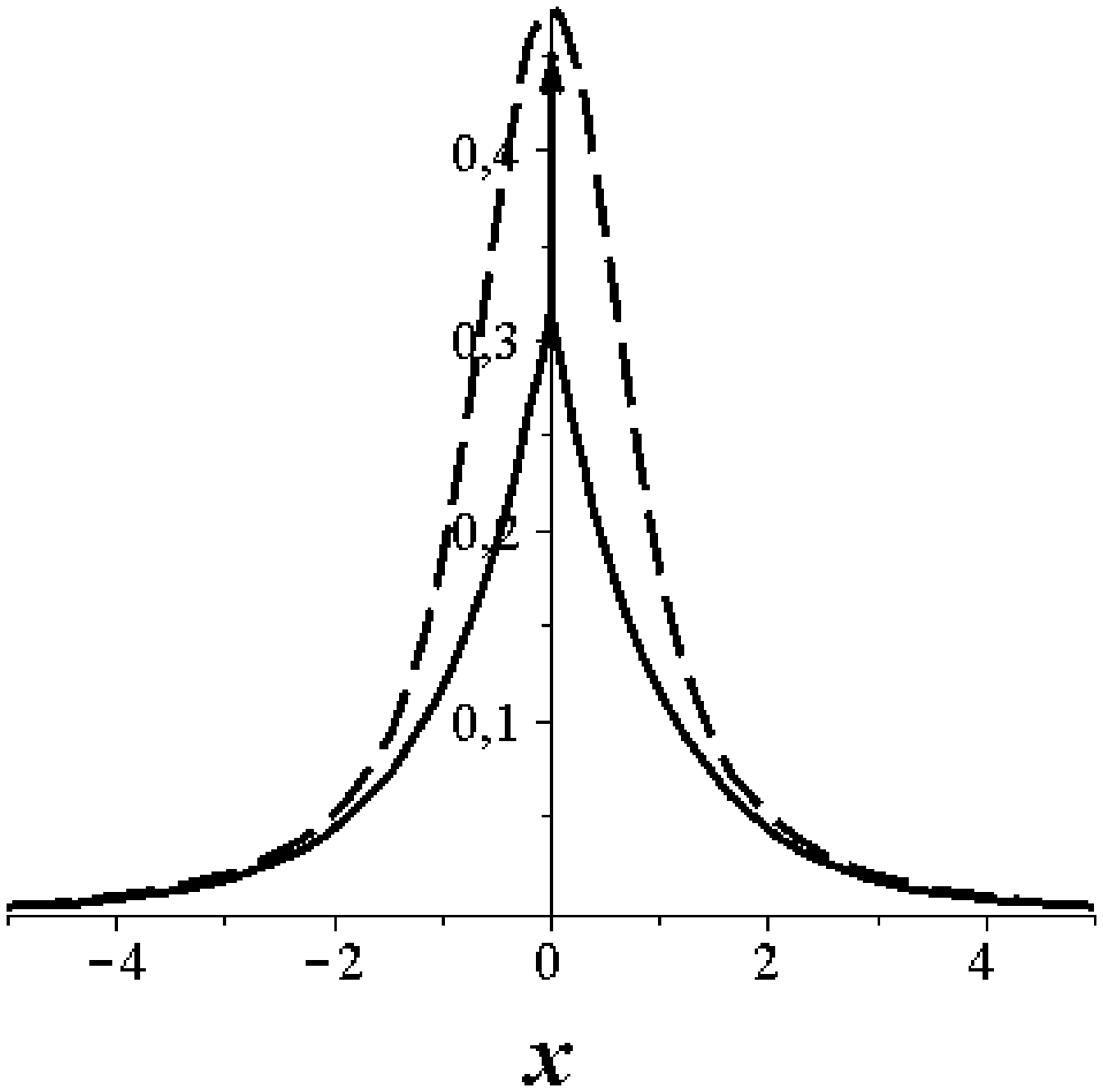}
\end{minipage}
\hspace{1.5cm}
\begin{minipage}{0.4\columnwidth}
\includegraphics[scale=0.35]{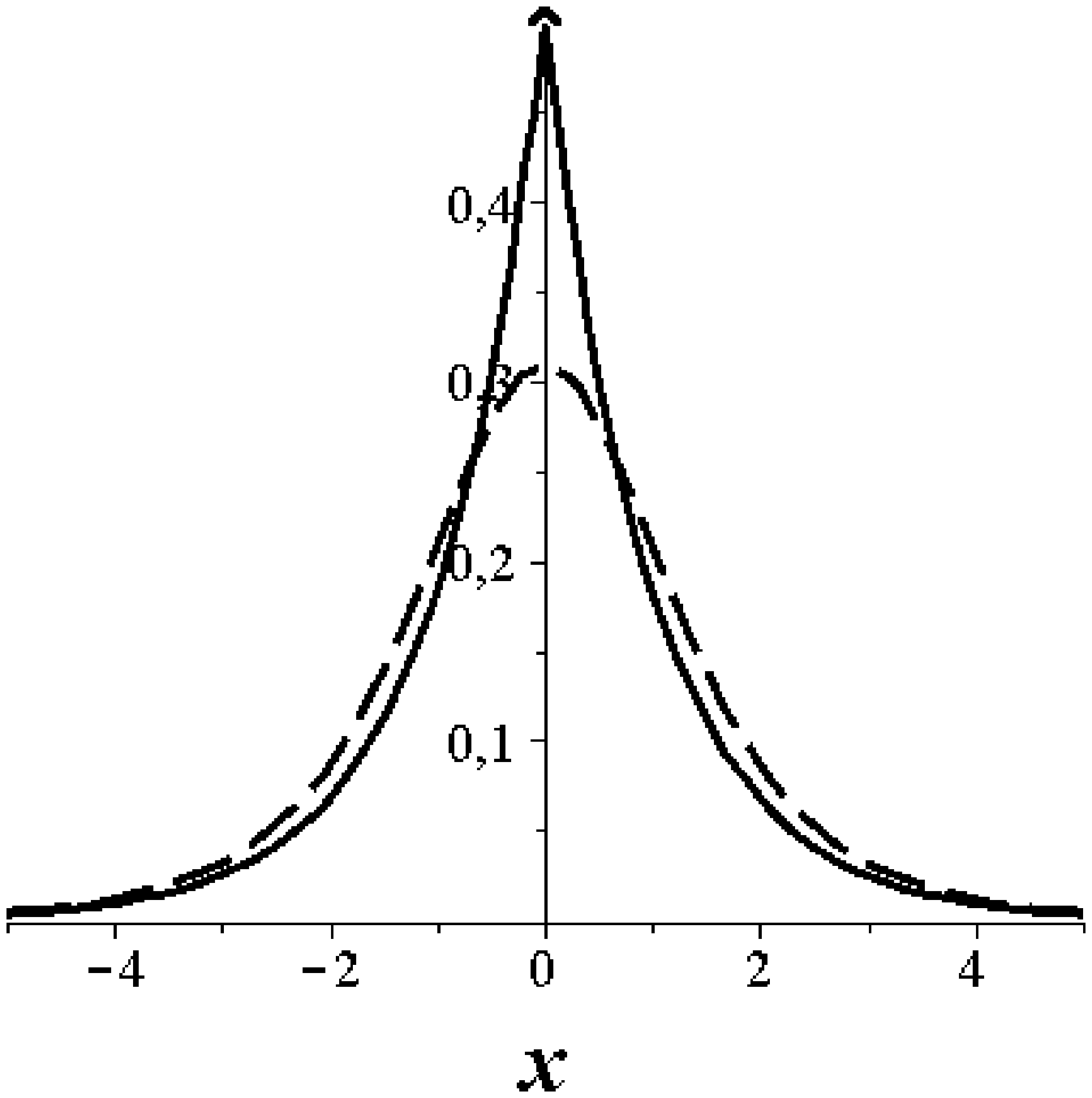}
\end{minipage}
\caption{ Fundamental solution for $k=1$, $\beta=1$, $n=1$ for $\sigma=0$ (solid) and $\sigma>0$ (dash). Left: $t=0.5$, right: $t=100$.}\label{Pic1}
\end{figure}

2. For $\sigma>0$ we have
\begin{eqnarray*}\label{E1s}
&&{\mathcal E}(t,x)=
\left[ 2 \cosh k x_1-{\rm Erf}\left(\frac { -2 \beta\,x_1+\sigma^2 k (1-  e^{-2\beta t})  }{2\sigma \sqrt {\beta}\sqrt {1- e^{-2 \beta t}}}\right)
e^{-{k x_1}} -\right.\\&&\left.
{\rm Erf}\left(\frac { 2 \beta\,x_1+\sigma^2 k (1-  e^{-2\beta t})  }{2\sigma \sqrt {\beta}\sqrt {1- e^{-2 \beta t}}}\right)
e^{k x_1} \right] \frac{k (1-e^{-2\beta t})}{4} e^\frac{k^2 \sigma^2 (1-e^{-2\beta t})}{4\beta} +\\&&
\frac{\sqrt{\beta} e^{-2\beta t} }{\sqrt {1- e^{-2 \beta t}}} e^{-\frac{\beta x_1^2}{\sigma^2 (1-e^{-2 \beta t})}},\\
&&x_1=x-\frac{B}{\beta}(1-e^{-\beta t}),
\end{eqnarray*}
 the limit as $t\to \infty$ is
\begin{eqnarray*}
&&{\mathcal E}_{st}(x)=
  \frac{k}{4}  e^{\frac{\sigma^2 k^2}{4\beta}} \left[2 \cosh k(x-\frac{B}{\beta}) -\right.
  \\&& \left. {\rm Erf}\left( -\frac{\beta x-B}{\sigma} +\frac{\sigma k}{2\sqrt{\beta}}\right) e^{-k(x-\frac{B}{\beta})}- {\rm Erf}\left( \frac{\beta x-B}{\sigma} +\frac{\sigma k}{2\sqrt{\beta}}\right) e^{k(x-\frac{B}{\beta})}
\right].
\end{eqnarray*}
Fig.\ref{Pic1} shows the comparative behavior of  the fundamental solution for $n=1$ for the cases $\sigma=0$ and $\sigma>0$ at different times.

\begin{figure}[htb]
\begin{minipage}{0.4\columnwidth}
\includegraphics[scale=0.35]{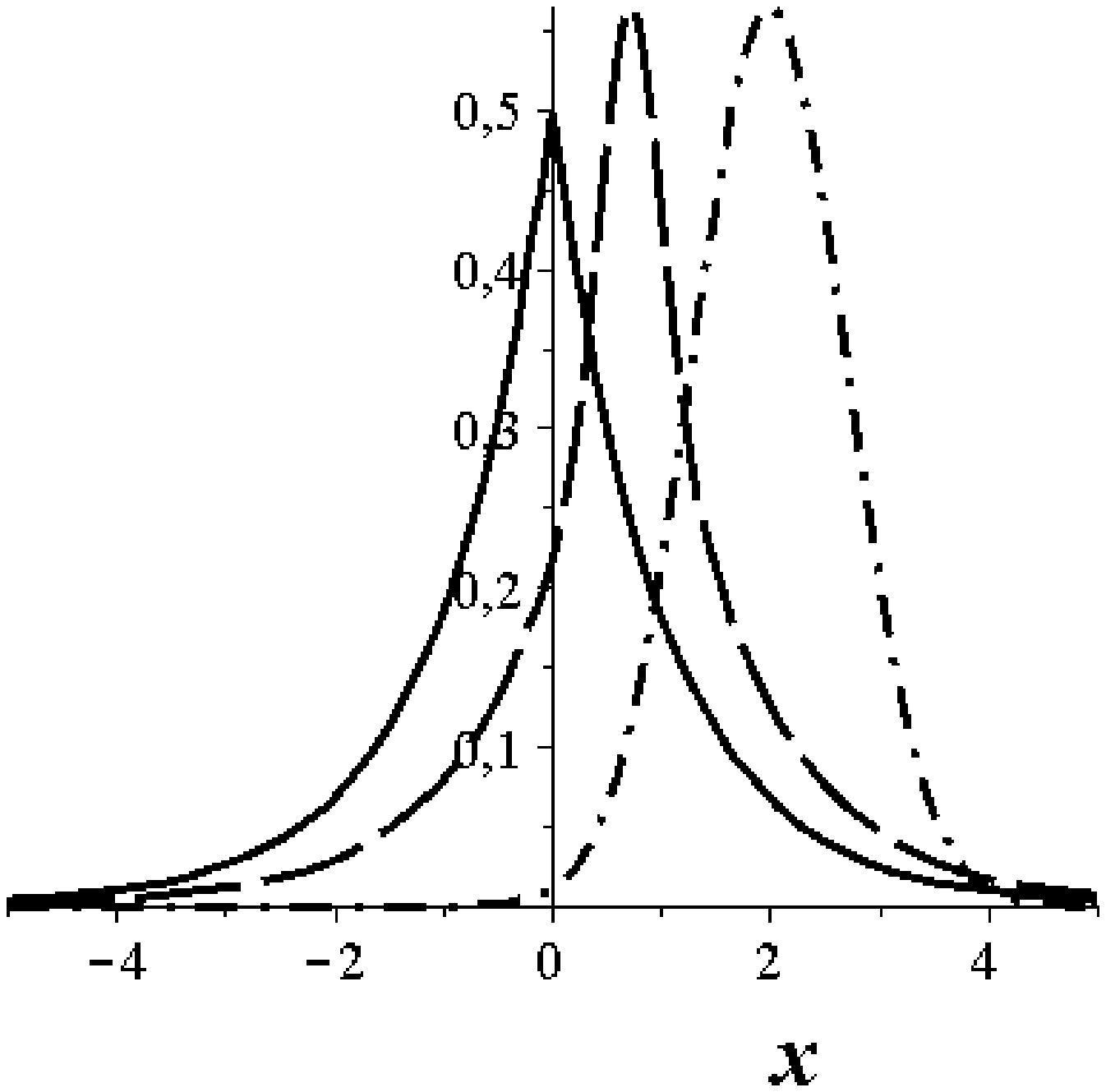}
\end{minipage}
\hspace{1.5cm}
\begin{minipage}{0.4\columnwidth}
\includegraphics[scale=0.35]{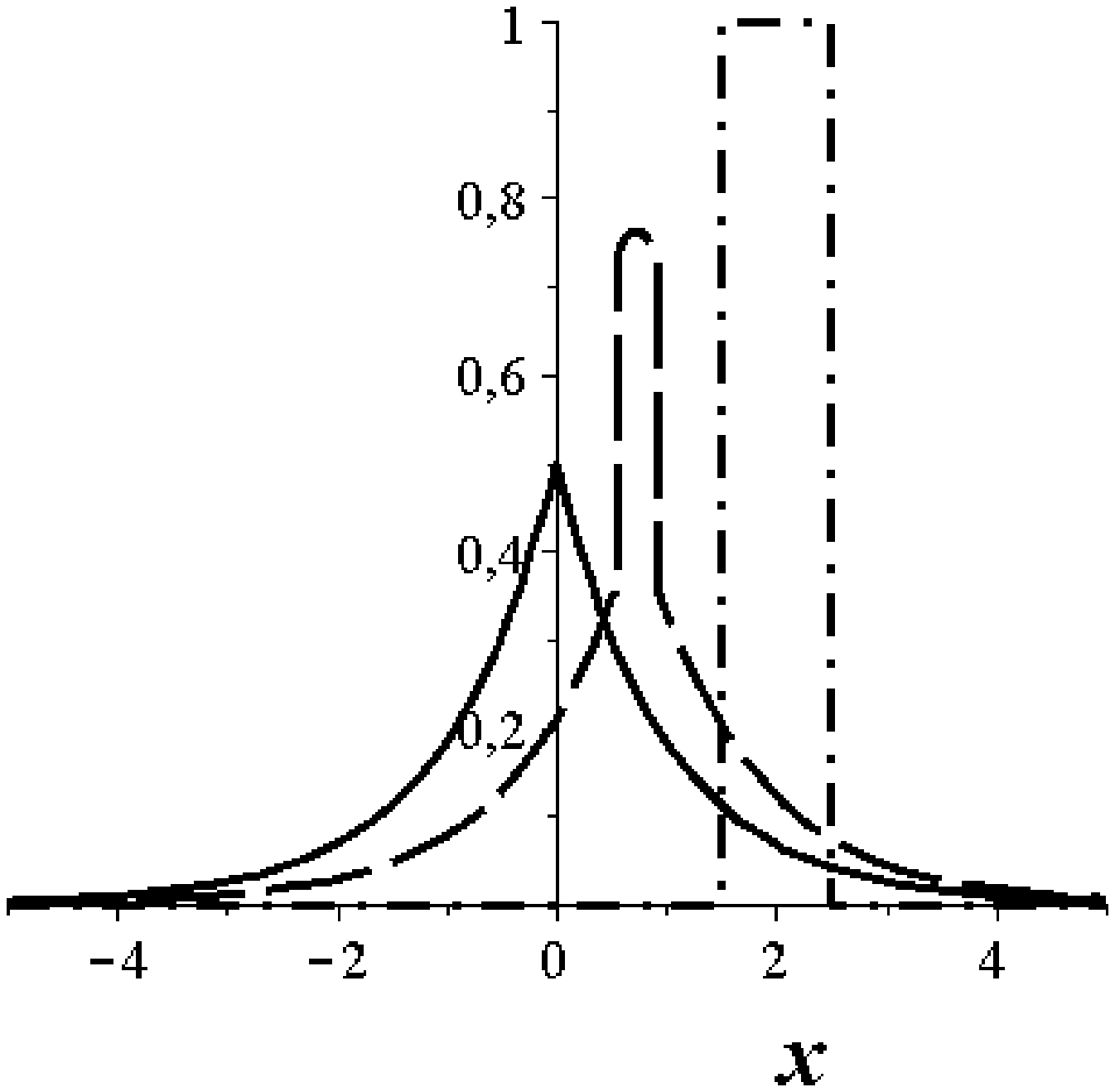}
\end{minipage}
\caption{ Dynamics of  density with initial data \eqref{gauss} (left) and \eqref{const} (right) at $a=2$, $k=1$, $\beta=1$, $\sigma=0$, $n=1$ at times $t=0$ (dash-dot), $t=1$ (dash), $t=10$ (solid). }\label{Pic2}
\end{figure}

3. Given some initial data, we can compute the convolution \eqref{conv} and obtain an explicit expression for the density $P(t,x)$. For example, this can be done with the initial Gaussian density distribution
\begin{equation}\label{gauss}
P|_{t=0}=\frac{1}{\sqrt{\pi}} e^{-(x-a)^2}.
\end{equation}
For $\sigma=0$
\begin{eqnarray*}
&&{P}(t,x)=   \frac{k(1-e^{-2\beta t})}{4}  e^{\frac{k^2 e^{-2\beta t} }{4}} ({\rm Erf} X -1) e^{-k|x-e^{-\beta t}a|}+\frac{e^{-\beta t}}{\sqrt{\pi}} e^{-(x-a e^{-\beta t})^2},\\
&&X=-x e^{\beta t} +\frac12 k e^{-\beta t}+a.
\end{eqnarray*}
It is easy to see that the limit of the density as $t\to \infty$ coincides with \eqref{P1}, and a weak discontinuity of the initially infinitely smooth solution exists for all $t>0$. Fig.\ref{Pic2}, left,  shows the evolution of the Gaussian density over time.

4. It is quite interesting to trace the evolution of a discontinuous density. Let us choose as initial data a piecewise constant
function
\begin{equation}\label{const}
P|_{t=0}=\Theta \left(x+a+\frac12\right)-\Theta \left(x+a-\frac12\right),\quad a=\rm const,
\end{equation}
where $\Theta $ is the Heaviside function.
The convolution \eqref{conv} can be also obtained explicitly. For   $\sigma=0$ it is
\begin{eqnarray*}
&&P(t,x)=C_0(t)+C_1(t)\Theta(\xi_+)+C_2(t) \Theta(\xi_-)+C_3(t)\Theta(\eta_+)+C_4(t) \Theta(\eta_-),\\
&&\xi_\pm=x +\left(a\pm\frac12\right) e^{-\beta t}, \quad \eta_\pm=x +\frac12 e^{-\beta t},
\end{eqnarray*}
we do not write down the time-dependent coefficients $C_i$, $i=0,\dots, 4.$
 Fig.\ref{Pic2}, right,  shows the evolution of the step density over time.
 Note that the jumps are preserved for all $t>0$, but their amplitude tends to zero for $t\to\infty$.

\begin{rem} In the case of $\sigma>0$ the density belongs to $C^\infty$ for every $t>0$ for all integrable data, even discontinuous ones, this follows from the fact that the fundamental solution belongs to $C^\infty $.
\end{rem}

\section{Discussion}

Note that other models  describe non-standard diffusion, in particular, in terms of fractional derivatives (see \cite{Metzler}, \cite{Lemaitre} for an exhaustive review and a detailed list of applications for which there is experimental evidence for the insufficiency of the usual Wiener process). For such models, in some cases, it is also possible to construct exact stationary density distributions, as a rule, expressed as special functions \cite{Metzler}. In particular, for the case of subdiffusion (slower than Gaussian), similar phenomena  to the presence of pure jumps also arise. This is, in particular, the non-smoothness of the density function.

The method we use can be generalized to the multidimensional case, including asymmetric diffusion. The method can also be modified for the case of anomalous diffusion by replacing the $\Delta$ operator with $-\sqrt{-\Delta}$ in the equation for the probability density \eqref{FPF}. However, the fundamental solution in this case is not as simple as above, it can be expressed in terms of exponential integrals.

\section{Author contributions}
Conceptualization,  methodology, writing, visualization, supervision, O.R.; investigation and validation,  O.R. and N.K.

\section{Acknowledgments}
O.Rozanova was supported  by the Russian Science Foundation under grant no. 23-11-00056,
performed at Рeoples’ Friendship University of Russia (RUDN University).
N.Krutov was supported by the Moscow Center for Fundamental and Applied Mathematics.
The authors are grateful to the anonymous reviewer for his/her careful reading.


\begin{thebibliography}{15}

  \bibitem{Billings}
 L. Billings, M.I. Dykman, I.B. Schwartz,
 Thermally activated switching in the presence of non-Gaussian noise,
Phys. Rev. E 78, 051122, 2008.

\bibitem{Chen}
X. Chen, Y.-M. Kang, Y.-X. Fu,
Switches in a genetic regulatory system under multiplicative non-Gaussian noise,
Journal of Theoretical Biology,
 435,
134-144, 2017.


\bibitem{Cont}
R. Cont, P. Tankov, Financial Modeling with Jump Processes. Chapman and
Hall, Boca Raton, 2004.

\bibitem{Denisov}
S. Denisov, W.Horsthemke, P.  H\"anggi, Generalized Fokker-Planck equation: Derivation and exact solutions, Eur. Phys. J. B 68, 567–575, 2009.

\bibitem{Ferreira}
M. Ferreira, N. Vieira, Fundamental solutions of the time fractional diffusion-wave and parabolic Dirac operators. J. Math. Anal. Appl. 2016 (447) 329–353.


\bibitem{Huang} 
G.-R. Huang,  D.B. Saakian, O.S. Rozanova, J.-L. Yu, C.-K. Hu,
Exact solution of master equation with Gaussian and compound Poisson noises,
J. Stat. Mech.  P11033, 2014.

\bibitem{Gardiner}
C. Gardiner, Stochastic Methods: A Handbook for the Natural and Social Sciences. Springer, 2009.

\bibitem{Borzi} B. Gaviraghi, M. Annunziato, A. Borzi, Analysis of splitting methods for solving a partial integro-differential Fokker–Planck equation, Applied Mathematics and Computation,
294, 1-17, 2017.

\bibitem{Piatnitski} A. Grigoryan, Y. Kondratiev, A. Piatnitski, E. Zhizhina,
Pointwise estimates for heat kernels of convolution-type operators,   Proceedings of the London Mathematical Society, 117 (4),
849-880, 2018.

\bibitem{Knopova}
V. Knopova, A. Kulik, Parametrix construction for certain Lévy-type processes, Random Operators and Stochastic Equations,  23 (2)  111-136, 2015. 



\bibitem{Kogan}
S. Kogan, Electronic Noise and Fluctuations in Solids 2nd edn (Cambridge: Cambridge University Press), 2008.


\bibitem{Kuhn}
F. K\"uhn, Transition probabilities of Lévy-type processes: Parametrix construction, Mathematische Nachrichten, 1–19, 2018.

\bibitem{Luchko}
Y. Luchko, On some new properties of the fundamental solution to the multi-dimensional space- and time-fractional diffusion-wave equation, Mathematics 2017, 5(4), 76.

\bibitem{Mainardi}
F.Mainardi, Y. Luchko, G. Pagnini,  The fundamental solution of the space-time fractional diffusion equation. Fract. Calc. Appl. Anal. 2001, 4, 153–192.

\bibitem{Maller}
 R.A. Maller, G. M\"uller, A. Szimayer, Ornstein–Uhlenbeck processes and extensions. In: Mikosch, T., Kreiß, JP., Davis, R., Andersen, T. (eds) Handbook of Financial Time Series. Springer, Berlin, Heidelberg, 2009.



\bibitem{Marko}
N.F. Marko,  R.J. Weil,  Non-Gaussian distributions affect identification of expression patterns, functional annotation, and prospective classification in human cancer genomes. PLoS ONE 7(10), e46935, 2012.

\bibitem{Metzler}
R. Metzler, J. Klafter,  The random walk's guide
to anomalous diffusion:
a fractional dynamics approach, Physics Reports, 339, 1-77, 2000.

\bibitem{Lemaitre}
E. Lemaitre, I.M. Sokolov, R.Metzler, A.V. Chechkin, Non-Gaussian displacement distributions in models of
heterogeneous active particle dynamics, New J. Phys. 25, 013010, 2023.









\bibitem{Peszat}
S. Peszat, Lévy–Ornstein–Uhlenbeck transition semigroup as second quantized operator, Journal of Functional Analysis, 260, 12, (3457-3473), 2011.

\bibitem{Picard}
J. Picard, On the existence of smooth densities for jump processes, Probab.
Theory Related Fields, 105,  481-511, 1996.


\bibitem{Priola}
E. Priola, J. Zabczyk,
Densities for Ornstein-Uhlenbeck processes with jumps,
Bulletin of the London Mathematical Society 41(1), 2008.

\bibitem{Rudenko}
O.V. Rudenko, A.A. Dubkov, S. N. Gurbatov
On exact solutions to the Kolmogorov–Feller equation,
Doklady Mathematics, 94(1) 476-479, 2016.

\bibitem{Schuss}
Z. Schuss, Theory and Applications of Stochastic Processes: an Analytical Approach.
Springer, 2010.








\bibitem{Wang}
G. Wang, Y. Wu, F. Xiao, Z. Ye, Y. Jia,
Non-Gaussian noise and autapse-induced inverse stochastic resonance in bistable Izhikevich neural system under electromagnetic induction,
Physica A: Statistical Mechanics and its Applications,
598, 127274, 2022.




\bibitem{Yang}
 A. Yang,  H. Wang,  T. Zhang, S. Yuan,
Stochastic switches of eutrophication and oligotrophication: Modeling extreme weather via non-Gaussian Lévy noise,
Chaos, 32, 043116, 2022.



\end{thebibliography}
\end{document}